\documentclass[referee,useAMS,usenatbib]{mn2e}
\usepackage[utf8]{inputenc}
\usepackage{amsmath}
\usepackage{graphicx}
\usepackage{epsfig}
\usepackage{chngcntr}
\usepackage{soul}
\usepackage[font={small,it}]{caption}
\usepackage{textcomp}
\usepackage{subfig}
\def\nel{n_{{\rm e}^-}}
\def\np{n_{{\rm p}^+}}
\def\me{m_{{\rm e}^-}}
\def\mp{m_{{\rm p}^+}}
\def\roe{\rho_{{\rm e}^-}}
\def\tmat{T^{\mu \nu}_{\rm matter}}
\def\tem{T^{\mu \nu}_{\rm em}}

\def\ie{{\em i.e., }}
\def\vp{v_{\rm p}}
\def\vfi{v_{\phi}}
\def\bp{B_{\rm p}}
\def\bfi{B_{\phi}}

\def\tb{\textbf}
\def\Psia{\Psi_{A}}
\def\psia{\psi_{A}}
\def\msin{\mbox{sin}}
\def\mcos{\mbox{cos}}
\def\mtan{\mbox{tan}}
\def\m2{M^{2}}
\def\g2{G^{2}}
\def\x2{x^{2}}
\def\xa2{x^{2}_{A}}
\def\4xa{x^{4}_{A}}
\def\f2{F^{2}}
\def\2mu{\mu^{2}}
\def\sigm{\sigma_{M}}
\def\sigmm{\sigma^{2}_{M}}
\def\siga{\sigma_{A}}
\def\thetaa{\theta_{A}}
\def\lsim{\lower.5ex\hbox{$\; \buildrel < \over \sim \;$}}
\def\gsim{\lower.5ex\hbox{$\; \buildrel > \over \sim \;$}}
\newcommand{\RN}[1]{
}

\usepackage{graphicx}
\title[Relativistic magnetized outflows]
{Study of relativistic magnetized outflows with relativistic equation of state}
\author[Singh \& Chattopadhyay]
{Kuldeep Singh$^{1,2}$, Indranil Chattopadhyay$^{1}$\thanks{Email:
kuldeep@aries.res.in (KS); indra@aries.res.in (IC)}\\
$^{1}$Aryabhatta Research Institute of Observational Sciences 
(ARIES), Manora Peak, Nainital-263002, India.\\
$^{2}$University of Delhi, Delhi, India.}
\begin{document}
\date{}
\maketitle
\label{firstpage}

\begin{abstract}
We study relativistic magnetized outflows using relativistic equation of state having 
variable adiabatic index ($\Gamma$) and composition parameter $(\xi)$. We 
study the outflow in special relativistic magneto-hydrodynamic regime, from sub-Alfv\'enic to super-fast domain.
We showed that, after the solution crosses the fast point, magnetic field collimates the flow and may form
a collimation-shock due to magnetic field pinching/squeezing. Such fast, collimated outflows may be considered as
astrophysical jets. Depending on parameters, the terminal Lorentz factors of an electron-proton outflow can comfortably exceed
few tens.
We showed that due to the 
transfer of angular momentum from the field to the matter, the azimuthal velocity of the outflow may flip sign. We also
study the effect of 
composition $(\xi)$ on such magnetized outflows. We showed that relativistic outflows are affected by the location
of the Alfv\'en point, the polar angle at the Alfv\'en point and also the angle subtended by the field lines with the equatorial plane,
but also on the composition of the flow. The pair dominated
flow experiences impressive acceleration and is hotter than electron proton flow.
\end{abstract}

\section{Introduction}
\label{sec:intro}
There are many observational evidences that astrophysical objects like young stellar
objects (YSOs), accreting white dwarfs, X-ray binaries (XRBs), and active galactic nuclei (AGN) produce jets.
AGN jet is a relativistic, collimated 
outflow which spans over large distance (few kpc to Mpc scale) with Lorentz factor ($\gamma$) ranging from few to few tens.
There are
mainly three things which are common in astrophysical jets --- first, jets are collimated \citep{asa12}, second, jets propagate with
high speeds \citep{p81} and the third is the over-collimation of the flow due to interaction with the ambient medium and/or by the
magnetic field pinching \citep{asa12,lrl12}. 
The bright knots observed throughout the jet,
may occur due to the existence of
multiple shocks caused by magnetic pinching, or, interaction with the ambient medium. There are many processes that can drive the jet outward with significant speed.
In principle, thermal-pressure gradient term can accelerate the jet to speeds comparable to the sound speed
at the jet base \citep{lckhr16}. However, \citet{mstv04} used the thermal-pressure gradient term as the main accelerating
process,
but achieved fast outflows by tweaking the equation of state of the flow.
The intense radiation field emanating from the associated accretion disc, may transfer momentum or energy to the jet material
and thereby accelerate it \citep[see,][]{ftrt85,cc00,pk04,c05,vkmc15,vc17,vc18,vc19}. It may be noted that, only a luminous disc
can radiatively drive a powerful jet, which would preclude possibility of powerful jets associated with under luminous accretion discs.
Therefore, the scientific community believes that the magnetic driving is a more general physical process, through which
powerful jets can be produced both in microquasars and AGNs. 

Global magnetized outflow solutions, i.e., solutions connecting the base of the outflow with the asymptotically large distance,
were first obtained by \citet{wd67}, which crossed the critical points (slow, Alfv\'en and fast) smoothly,
albeit on the equatorial plane. \cite{wd67} model predicted the correct wind speed at the earth orbit.
In the cold flow regime, \citet{bp82} proposed a model of centrifugally and magnetically driven outflow
from cold Keplerian disc, somewhat like a bead flung by a rotating wire. A novel idea as it was, but the cold flow assumption
limited its applicability in studying outflows launched from the hot inner regions of accretion discs around compact objects.
\citet{lmms86} developed the magneto-hydrodynamic (MHD) equations of motion for accretion channel on to strongly magnetized compact stars
and was later used to study accretion on to neutron stars and white dwarfs \citep{kkm08,sc18}.
Streamline of a magnetically driven outflow should originate from the accretion disc on the equatorial
plane, but as the plasma flows out, the streamline should move away from the equatorial plane and around the rotation axis.
Indeed, there were few papers which showed that open field lines, coming from the underlying disc, collimate the jet around
the rotation axis \citep{sa85, sa87, lbc91}. 
But most of these models were either in the non-relativistic regime, or, in the cold plasma regime, or both. 

\citet{li92} extended these cold flow to
relativistic regime and studied the radially self-similar jet solutions. Then, \cite{vla03a,vla03b} further extended
the cold relativistic MHD to
hot flow by including the thermal-pressure gradient term. Therefore, outflows with
relativistic bulk speed and temperature, could be studied. The thermal-pressure gradient term dominates near the jet base and
can accelerate 
the flow near the base, but it is unlikely to do so at larger distance away from the jet-base.
\citet{pol10} used \citet{vla03a} model with fixed adiabatic index
($\Gamma=5/3$) 
equation of state and showed that the flow can become trans-Alfv\'enic (sub Alfv\'enic to super Alfv\'enic)
and trans-fast (sub fast to super fast). In contrast, \citet{vla03a}
could obtain only trans-Alfv\'en flow with $\Gamma=4/3$. Therefore, the thermodynamics of the flow may play an important role
in determining the nature of the solution. 
In particular, the outflow is hot near the base but the temperature decreases by few orders of magnitude at large distances,
therefore the adiabatic index is not likely to remain constant through out the flow. 

In this paper, we obtain radially self-similar solutions of magnetically driven relativistic outflows by following the
methodology of \citet{pol10},
but instead of using a fixed $\Gamma$ EoS, we consider a relativistic EoS.
We use a relativistic EoS that was proposed by \citet{cr09}, which was inspired by earlier works \citep{c38,s57,cg68,rcc06}.
We would like to find out, whether we still get trans-Alfv\'enic, trans-fast flow with an EoS which has no fixed value of $\Gamma$.
We focus on how the jet solutions changes with the change in current distribution, Alfv\'en point, Alfv\'en point polar angle and other
flow parameters.
We compare an outflow solution described by relativistic EoS, with the one described by
fixed $\Gamma$ EoS \citep{vla03a,vla03b,pol10}. An interesting aspect would be to study and compare flows with
different plasma composition parameter. As far as we know, such an effort has not been considered for relativistic MHD outflows.
In short, we would like to investigate how would various flow parameters affect the magnetically driven relativistic
outflow.


The order of the paper is as follows, in section \ref{subsec:govereqs}, we present special relativistic MHD equations.  
In section \ref{subsec:clo}, we discuss the two closure equations, one is flux freezing condition and other is
the relativistic EoS having variable adiabatic index. Reduced relativistic MHD equations are presented in section \ref{subsec:mhdeqs}. 
Methodology to solve equations of motion are explained in section \ref{sec:meth}. In section \ref{sec:result} we present the results of
outflow solutions. Discussions and concluding remarks are presented in section \ref{sec:conclude}.

\section{Relativistic MHD equations and assumptions}
\subsection{Governing equations}
\label{subsec:govereqs}
Equations of motion of relativistic magneto-hydrodynamics (RMHD) can be obtained from the four divergence of
the total energy-momentum tensor. The
energy-momentum tensor for matter is, $\tmat=(\bar{e}+p)u^{\mu}u^{\nu}+p\eta^{\mu \nu}$, where
$\bar{e}$ is energy density, $p$ is gas pressure, the four-velocity component $u^{\mu}=
\left(\gamma c, \gamma \tb{v}\right)$, $\eta^{\mu \nu}=\mbox{diag}\left[-1,1,1,1\right]$ and
$c$ is the speed of light. The energy-momentum tensor of the electromagnetic field is given by
$\tem=\left(F^{\mu \lambda}F^{\nu}_{\lambda} - \frac{1}{4} \eta^{\mu \nu}F^{\delta \lambda}F_{\delta \lambda}\right)/(4 \pi)$. Therefore, the total energy-momentum tensor
is $T^{\mu \nu} = \tmat + \tem$. The conservation
of energy and momentum in a covariant form can be written as,
\begin{equation}
\nabla_{\nu}T^{\mu \nu} = 0
{\label{cormhd.eq}}
\end{equation}

Maxwell's equations are, 
\begin{equation}
\nabla . \tb{B}=0,~~ \nabla . \tb{E}=\frac{4\pi}{c} J^{0},~~
\nabla \times \tb{B}=\frac{1}{c}\frac{\partial \tb{E}}{\partial t} + \frac{4\pi}{c}\tb{J},~~
\nabla \times \tb{E}=-\frac{1}{c}\frac{\partial \tb{B}}{\partial t}, 
{\label{maxs.eq}}
\end{equation}
where $J^{\mu}=\left(J^{0},\tb{J}\right)$ is the four-current. 

\subsection{Closure equations}
\label{subsec:clo}
To solve the above set of equations (\ref{cormhd.eq} and \ref{maxs.eq}) we need two more 
equations, because the number of variables are more than the number of equations. For matter, we 
need an equation which relates the thermodynamic variables \ie EoS of the fluid. We also need another equation which relates
the electric field to the magnetic field.

\subsubsection{Relativistic EoS having variable $\Gamma$}
\label{subsec:eos} 
In this study we have used relativistic EoS for multi-species flow which was proposed by
\citet[][ also called as CR EoS]{cr09}, which is 
given by,
\begin{equation}
\bar{e} = \nel \me c^{2}f(\Theta,\xi) = \roe c^{2}f(\Theta,\xi) = \frac{\rho c^{2}f(\Theta,\xi)}{K},
\label{etrnl.eq}
\end{equation} where, $K = [2-\xi(1-1/\eta)]$, $f(\Theta,\xi) = (2-\xi)\left[1 + \Theta\left(\frac{9\Theta + 3}
{3\Theta + 2}\right)\right] + \xi\left[\frac{1}{\eta}
+ \Theta\left(\frac{9\Theta + 3/\eta}{3\Theta + 2/\eta}\right)\right]$, $\Theta={\kappa_{\rm \small B}T}/{\me c^{2}}
$ is the dimensionless temperature, $\roe$ is the rest-mass density of electrons, $\rho$ is the rest-mass 
density, $\eta={\me/\mp}$ is electron to proton mass ratio, the composition parameter $\xi={\np/\nel}
$ is the ratio of number density of protons to that of electrons.
A flow described by $\xi=0.0$ implies an electron-positron pair plasma, $0.0<\xi<1.0$ imply electron-positron-proton plasma and 
$\xi=1.0$ implies electron-proton plasma. 
Enthalpy $h$, variable adiabatic index $\Gamma$ and sound speed $c_{\rm s}$ are given by,
\begin{equation}
h = \frac{\bar{e} + p}{\rho} = \frac{fc^{2}}{K} + \frac{2\Theta c^{2}}{K},
\label{enthp.eq}
\end{equation} and 
\begin{equation}
\Gamma = 1 + \frac{1}{N}, ~
N = \frac{1}{2}\frac{df}{d\Theta} \mbox{ and } c^{2}_{\rm s}=\frac{2\Theta \Gamma c^{2}}{f+2\Theta}.
{\label{gama.eq}}
\end{equation}
Integrating $1^{st}$ law of thermodynamics ($u_\mu \nabla_\nu T^{\mu \nu}=0$) with the help of continuity equation, we can obtain the adiabatic equation of state \citep{kscc13, vkmc15},
\begin{equation}{\label{rel_rho.eq}}
\rho={\cal K}g(\Theta,\xi),
\end{equation}
where, $g(\Theta,\xi)=\mbox{exp}(k_3) \Theta^{3/2}(3\Theta+2)^{k_1}(3\Theta+2/\eta)^{k_2}$, $k_1=3(2-\xi)/4$, $k_2=3\xi/4$ and $k_3=(f-K)/(2\Theta)$ and ${\cal K}$ is the measure of entropy. 
Therefore, pressure $p$ is given by,
\begin{equation}{\label{press.eq}}
p=\frac{2{\cal K}g(\Theta,\xi)\Theta}{K}c^{2}  
\end{equation}  

\subsubsection{Ideal MHD flow assumption}
\label{subsec:ideal}
For the ideal MHD flow, the electric field is zero in the co-moving frame \ie 
$u_{\nu}F^{\mu \nu}=0$ or 
\begin{equation}
\tb{E}=-\frac{1}{c}\tb{v}\times\tb{B}.
\label{elecf.eq}
\end{equation}
This is known as the ideal MHD condition. The flux freezing condition is obtained from the
Faraday equation,
\begin{equation}
\nabla\times(\tb{v}\times\tb{B})=\frac{\partial \tb{B}}{\partial t}
\end{equation}

\subsection{Conventional Relativistic MHD equations}
\label{subsec:mhdeqs}
By using the EoS and ideal MHD assumption, we can write equations (\ref{cormhd.eq}) and (\ref{maxs.eq}) in the 
conventional form.

The mass conservation equation is $\nabla_{\mu}(\rho u^{\mu})=0$, or the continuity equation,
\begin{equation}
\frac{\partial\left(\gamma\rho\right)}{\partial t} + \nabla.\left(\tb{v}\gamma\rho\right)=0.
\label{mascon.eq}
\end{equation}

The momentum conservation equation is, $\nabla_{\nu}T^{k\nu}=0$, where the $k=1,2,3$ 
components,
\begin{equation}
\gamma\rho\left(\frac{\partial}{\partial t} + \tb{v}.\nabla\right)\left(h\gamma\tb{v}\right)=-\nabla p + \frac{J^{0}\tb{E} + \tb{J}\times\tb{B}}{c}.
\label{momcon.eq}
\end{equation} 

The first law of thermodynamics is obtained by going to the co-moving frame of the flow, $u_{\mu}T^{\mu \nu}_{,\nu}=0$,
\begin{equation}
\left(\frac{\partial}{\partial t} + \tb{v}.\nabla\right)e + p\left(\frac{\partial}{\partial t} + \tb{v}.\nabla\right)\left(\frac{1}{\rho}\right)=0,
\label{1stlaw.eq}
\end{equation} 
where $e \equiv \bar{e}/\rho$.\\
We study the axis-symmetric steady flow, therefore, $\partial / \partial t = 0$ and
$\partial / \partial \phi = 0$. For axis-symmetric flow, the solenoidal condition can be written as,
\begin{equation}
\nabla . \tb{B}=\nabla . {\bf \bp} = 0. 
\end{equation}
The total magnetic field ${\bf B}$ is given as, 
\begin{equation}
\tb{B}={\bf \bp}+{\bf \bfi},~~\mbox{where, } {\bf \bp}=\frac{\nabla A\times\hat{\tb{$\phi$}}}{\varpi}.
\end{equation} 
Here, ${\bf \bp}$ and $\bf \bfi$ are the poloidal and azimuthal components of the magnetic field, respectively.
The $A(\varpi,z)$ is a poloidal magnetic flux function and this can be defined as
$A=\frac{1}{2\pi}\int\int {\bf \bp}.d\tb{S}$
and ${\bf \bp}.\nabla A=0$ which means that poloidal magnetic field lines are orthogonal to the gradient of 
magnetic flux function. Here, $\varpi$ represents the cylindrical radius.
With the help of ideal MHD flow condition (\ref{elecf.eq}) and $E_{\phi}=0$
(from Faraday equation \ref{maxs.eq}) we can show that ${\bf \vp} \parallel \tb{B}_{\rm p}$, so
\begin{equation}
E=\frac{\varpi \Omega}{c}\tb{B}\times\tb{e}_{\phi}
\mbox{,~~}
\tb{v}=\frac{\Psia}{4\pi \gamma \rho}\tb{B} + \varpi \Omega \tb{e}_{\phi}
\mbox{ and } \frac{\Psia}{4\pi \gamma \rho}=\frac{\vp}{\bp}.
\label{defvel.eq}
\end{equation}
Here, $\Psia$ is the mass to magnetic flux ratio and $\Omega$ is the angular velocity of fieldlines.
We can obtain the constants of motion by projecting equations (\ref{mascon.eq})
- (\ref{1stlaw.eq}) along and perpendicular to the poloidal fieldlines and
then integrating them \citep[for more details see][]{vla03a, vla03b}, \footnote{for non-relativistic MHD, see \citet{h78}}
we have five constants of motion $\Omega(A),~\Psia(A),~L(A),~\mu(A)
,~{\cal K}(A)$. 
The poloidal Alfv\'enic Mach number \citep[see,][]{mi69} is defined as,
\begin{equation}
\nonumber
M\equiv \frac{\gamma v_{p}}{\left({{B_{p}}/{\sqrt{4\pi\rho h}}}\right)},
\end{equation}
and using equations (\ref{enthp.eq}), (\ref{press.eq}) and (\ref{defvel.eq}), we can also write $M$ as,
\begin{equation}
\m2=q(A)\frac{h(h-f(\Theta,\xi)/K)K}{2\Theta g(\Theta,\xi)}=q(A)\frac{h}{g(\Theta,\xi)},
\label{amach.eq}
\end{equation}
where $q(A)\equiv {\Psia^2}/{4\pi{\cal K}}$. 
To solve RMHD equations we assume that jet solutions are radially self-similar \citep[for more details see 
section 3 in][]{vla03a}.
The derivatives of dimensionless temperature $\Theta$ and
enthalpy $h$ w.r.t polar angle $\theta$ are given by, 
\begin{equation}
\frac{d\Theta}{d\theta}=-\frac{g(\Theta,\xi)\Theta K}{qN\left(hK-2\Gamma\Theta \right)}\frac{d\m2}{d\theta}
\mbox{ and }
\frac{dh^{2}}{d\theta}=-\left(\frac{2h^{2}}{\m2}\right)\frac{2\Gamma\Theta}{hK-2\Gamma\Theta}\frac{d\m2}{d\theta} 
\label{dTdh.eq}
\end{equation}
If we take the derivative of total energy \textit{w.r.t} polar angle $(\theta)$ with the help of equations
(\ref{amach.eq}) and (\ref{dTdh.eq}) we obtain \citep[for more details see appendix and][]{pol10},
\begin{equation}
A_{1}(\theta,\psi,G^{2},\m2)\frac{d\m2}{d\theta}+B_{1}(\theta,\psi,G^{2},\m2)\frac{d\psi}{d\theta}=C_{1}(\theta,\psi,G^{2},\m2),
\label{dengy.eq}
\end{equation}
where $x\equiv \varpi\Omega/c$ is cylindrical radius in terms of light-cylinder,
$G\equiv x/x_{A}$ (here, $x_A\equiv x$ at Alfv\'en point) and $\psi$ is the angle of poloidal field line with the disk.
The transfield equation which controls the collimation of the flow, can be obtained from the momentum equation 
by taking dot product with $-\nabla A$ \ie perpendicular to the poloidal field line, 
\begin{equation}
A_{2}(\theta,\psi,G^{2},\m2)\frac{d\m2}{d\theta}+B_{2}(\theta,\psi,G^{2},\m2)\frac{d\psi}{d\theta}=C_{2}(\theta,\psi,G^{2},\m2).
\label{trans.eq}
\end{equation}
Therefore, we can get the wind equation or outflow equation $({d\m2}/{d\theta})$ for radially self-similar
flows by solving equations (\ref{dengy.eq}) and (\ref{trans.eq}),  
\begin{equation}
\frac{d\m2}{d\theta}=\frac{C_{1}B_{2}-C_{2}B_{1}}{A_{1}B_{2}-A_{2}B_{1}}.
\label{dm2.eq}
\end{equation}

\section{Methodology}
\label{sec:meth}
We study 
the flow in special relativistic domain, in which the slow magnetosonic point does not form,
\ie we find the solution from the sub-Alfv\'enic to super-fast regime. 
To obtain the solution of magnetically driven relativistic outflow about the axis of symmetry, we
integrate equations (\ref{dTdh.eq})\footnote{Equation (\ref{amach.eq}) instead of equation
(\ref{dTdh.eq}) may also be used, since they are equivalent.} and (\ref{dm2.eq}).
In addition, we also solve
equation (\ref{dg.eq}) and total energy to mass flux 
ratio equation (\ref{ber.eq}) to obtain $\psi$ if the value of $\mu$ is known. 
First, we supply the values of Alfv\'en point $x_A$, $F$ (current distribution), $q$, $\theta_A=\theta|_{x_A}$, $\psia=\psi|_{x_A}$.
We obtain $ M^{2}_A~(=1-x^{2}_{A})$ and therefore $\Theta_A$ using equations (\ref{enthp.eq} \& \ref{amach.eq}).
Then we obtain $\sigma_A$ from equations (\ref{bera.eq} \& \ref{arc.eq}) for a given value of $\sigm$.
Now we obtain the value of $\mu$ and $p_A=d\m2/d\theta|_{x_A}$ from equations (\ref{bera.eq}) and
(\ref{sigA.eq}),
respectively.
With these values we integrate equations (\ref{dm2.eq}, \ref{dg.eq}, \ref{amach.eq} or \ref{dTdh.eq}) starting from $x_A$ inward and outward. The solution may not pass through the fast point, so we iterate on $\sigm$ until
the solution passes through the fast point as well. We use Runge-Kutta fourth order method to integrate but also use Newton-Raphson's method to accurately obtain the flow quantities like $\theta_{\rm f},~\psi_{\rm f},~\g2_{\rm f},\m2_{\rm f}$, where the suffix `f' denotes quantities measured at the fast-point.
Since, we integrate the equations starting from the Alfv\'en point, therefore
$\xa2,\thetaa,\psia$ essentially are the boundary conditions or boundary parameters. 
In the present paper, there is no need to specify adiabatic index $\Gamma$ since it is self-consistently
obtained from EoS. 
In addition to this, we have one more free parameter $\xi$ which controls the composition of the flow.


\section{Results}
\label{sec:result}
 \vspace{0.0cm}
\begin{figure}
\hspace{2.0cm}
\includegraphics[width=12cm,height=12cm]{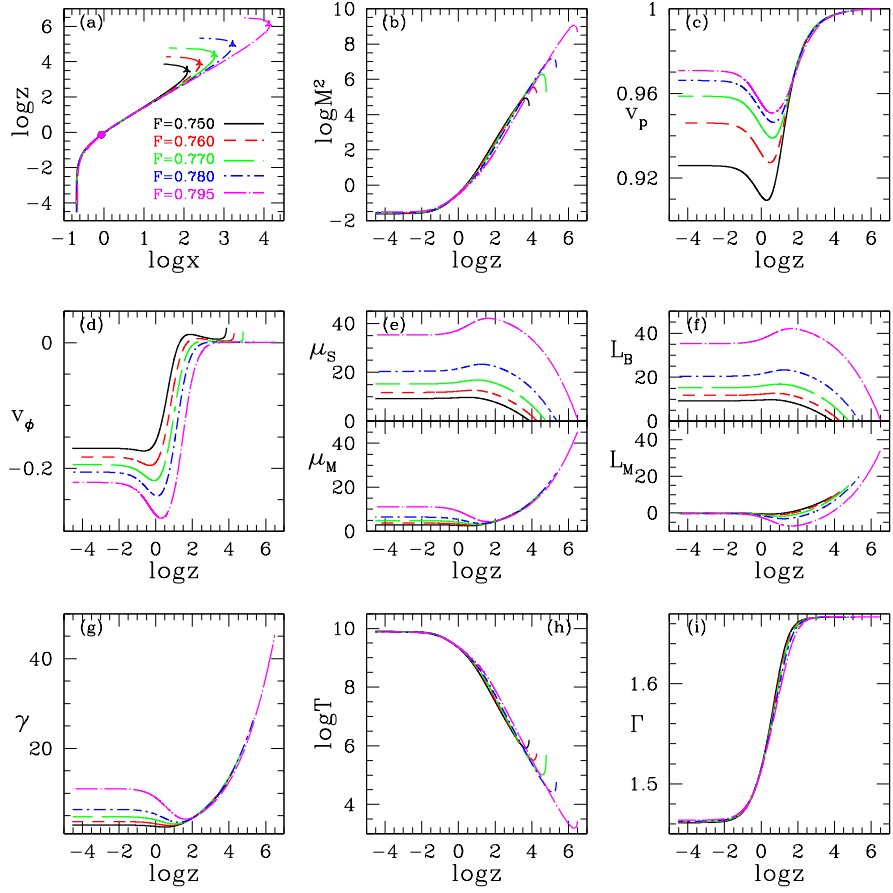} 
\caption{{Outflow solutions for different values of $F=0.750(\mbox{solid black})$,
$0.760(\mbox{dashed red})$,
$0.770(\mbox{long-dashed green})$, $0.780(\mbox{dashed-dotted blue})$, $0.795(\mbox{long-dashed-dotted magenta})$ 
and other four parameters are fixed \ie $x^{2}_{A}=0.75,\theta_{A}=50,\psia=55,q=500,\xi=1$. (a) Projected stream
line, (b) $log(M^{2})$, (c) $\vp$, (d) $\vfi$, (e) $\mu_{S}=-\varpi\Omega\bfi/\Psia c^{2}$ and
$\mu_{M}=\gamma h$, (f) $L_{B}$ and $L_{M}$, (g) $\gamma$, (h) $log(T)$ and (i) $\Gamma$, are plotted
with $log(z)$. Here, $z$ is vertical height and $x$ is cylindrical 
radius in units of light cylinder. Solid circles and triangles represent Alfv\'en point and fast-point locations.}}
\label{lab:fig1}
\end{figure}

In this paper, the velocity is measured in the units of speed of light $c$ and distance is in units of light cylinder
$r_{c}\equiv\frac{c}{\Omega}$. 
In our model, there are two main free input parameters $F$ and $q$, three boundary parameters $\psia,\thetaa,x^{2}_{A}$ and
a composition parameter $\xi$. We study the effect of these parameters on the outflow solutions and on the
collimation of outflowing matter with relativistic EoS.

\subsection{Solutions for different current distributions ($F$)}
In Fig.\ref{lab:fig1}, we plot different solutions for different current distribution parameter
$F=0.750$ (solid black), $0.760~(\mbox{dashed red})$, $0.770~(\mbox{long-dashed green})$, $0.780~(\mbox{dashed-dotted blue})$, $0.795~(\mbox{long-dashed-dotted magenta})$ 
and other four parameters are fixed \ie $x^{2}_{A}=0.75,~\theta_{A}=50,~\psia=55,~q=500$ \& $\xi=1.0$.
In Fig.\ref{lab:fig1}a, projected stream line in the $x-z$ plane is plotted. The distribution of corresponding flow variables like $log(M^{2})$ (Fig.\ref{lab:fig1}b), poloidal velocity
$\vp$ (Fig.\ref{lab:fig1}c), azimuthal velocity $\vfi$ (Fig.\ref{lab:fig1}d), Poynting to mass flux ratio $\mu_{S}\equiv-\varpi\Omega\bfi/\Psia c^{2}$ and
matter to mass flux ratio $\mu_{M}\equiv\gamma h$ (Fig.\ref{lab:fig1}e), angular momentum associated with the magnetic field
$L_{B}\equiv-\varpi\bfi/\Psia$ and matter $L_{M}\equiv h\gamma\varpi\vfi$ (Fig.\ref{lab:fig1}f),
Lorentz factor $\gamma$ (Fig.\ref{lab:fig1}g), $log(T)$ (Fig.\ref{lab:fig1}h) and adiabatic index $\Gamma$ (Fig.\ref{lab:fig1}i) with $log(z)$ are plotted. In Fig.\ref{lab:fig1}(a), solid-circles represent 
Alfv\'en point location and solid-triangles represent the fast point location, where $z$ is the vertical height and
$x$ is the cylindrical radius. In Fig.\ref{lab:fig1}(a), we note that if we 
increase $F$, the solution collimates at higher height $(z)$. 
Higher value of $F$ implies
weaker magnetic field near the base, so it travels larger $z$ before the outflow starts to collimate.
In panel Fig.\ref{lab:fig1}(c), we see that $\vp$ has a dip, which is due to the interaction of magnetic field
with matter. Near the base, $\mu_{S}$ gains at the cost of $\mu_{M}$ (Fig.\ref{lab:fig1}e), therefore there is
simultaneous decrease in thermal and kinetic terms.
When the magnetic energy ($\mu_{S}$) becomes sufficiently strong, it starts to accelerate the outflow, although
the outflow temperature continue to decrease. Hence there is a dip in $\vp$. Another very interesting 
result is that $\vfi$ changes sign from negative to positive (Fig.\ref{lab:fig1}d).
It means, initially the flow
is rotating clockwise and somewhere in between the Alfv\'en and the fast points, the flow flips to a counter-clockwise direction. In MHD, we have two types of
angular momentum, one that is associated with the matter $L_{M}\equiv h\gamma\varpi\vfi$ and the other associated with
the magnetic field $L_{B}\equiv-\varpi\bfi/\Psia$. Therefore, only total angular momentum is conserved throughout the
flow but not the individual angular momenta (Fig.\ref{lab:fig1}f). Thus, azimuthal velocity $\vfi$ changes sign
because of transfer of angular momentum from magnetic field to matter. In Fig.\ref{lab:fig1}(g), the 
variation of Lorentz factor $\gamma$ is shown. We can see that higher value of $F$ produces outflows with higher Lorentz factor
$\gamma\sim 45$  ($F=0.795,\mbox{ long-dashed-dotted}$).
In Fig.\ref{lab:fig1}(h), we plot temperature variation of the outflow with height, for different values of
$F$ parameter. We can see that outflow starts with high temperature when it is sub-Alfv\'enic and temperature
drops to very small value when the flow becomes super-fast. Last panel
Fig.\ref{lab:fig1}(i) shows that the adiabatic index $\Gamma$ does not remain constant throughout the solution,
it varies from $\Gamma\sim ~1.44$ to $5/3$. It is well known that gases with non-relativistic
temperatures have $\Gamma=5/3$ or the polytropic index $N=3/2$. For gases with ultra-relativistic
temperatures, $N\rightarrow 3$ or $\Gamma \rightarrow 4/3$. It may be noted that, $N$ is the temperature gradient of the specific
energy of the gas
i.e., $\sim df/d\Theta$ (see, equation \ref{gama.eq}). For non-relativistic thermal speed (for $T\lsim 10^7$K, the energy density of
the gas (${\bar e}$)
is dominated by rest-mass energy, so $N$ (therefore $\Gamma$) remains constant ($\equiv 5/3$). But for higher temperatures,
the thermal speed becomes relativistic, therefore kinetic contribution becomes comparable to rest mass in ${\bar e}$, as a result
$N$ increases with rising $T$. But the upper limit of thermal speed is $c$, therefore for ultra-relativistic temperature,
the kinetic contribution of the gas particles into ${\bar e}$ of the gas becomes maximum and therefore
$N$ again becomes temperature independent,
where asymptotically $N\rightarrow 3$ (or, $\Gamma \rightarrow 4/3$).
For example, if the temperature of a gas is in between these two extremes ($10^7~{\rm K} < T < 10^{13}$K), then the thermal state
is described by $3/2\lsim N \lsim 3$
\citep[see, figure 1a of][]{cr09}.
In Fig.\ref{lab:fig1}(h), temperature drops from $\sim 10^{10}$ to $\sim 10^4$ the thermal energy decreases
as a result, $\Gamma$ changes from $\sim 1.44$ (near-relativistic) to $\sim 5/3$ (non-relativistic).
 
\begin{figure}
  \centering
  \subfloat[Streamline side view.]{\includegraphics[width=0.55\textwidth]{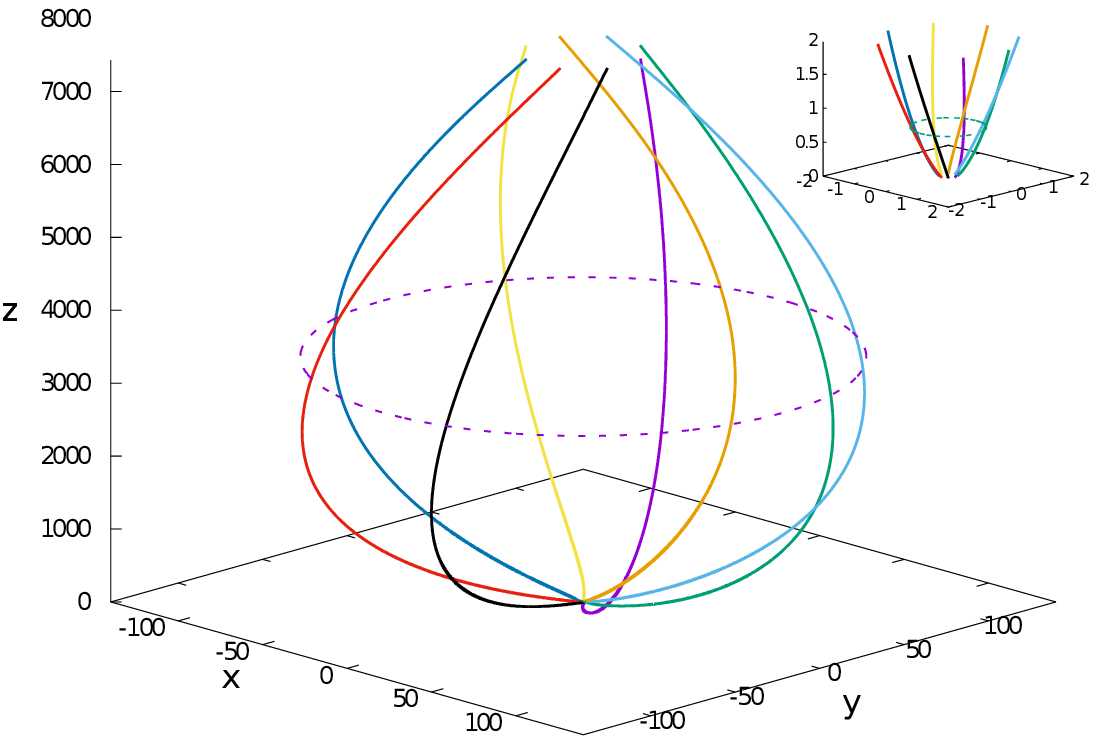}\label{fig:f2a}}
  \hfill
  \subfloat[Streamline top view.]{\includegraphics[width=0.38\textwidth]{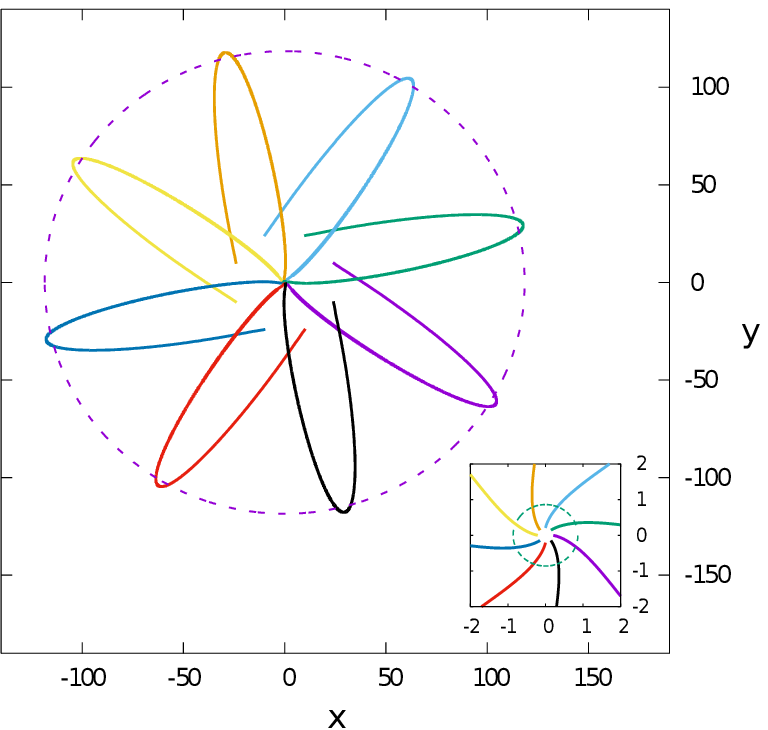}\label{fig:f2b}}
  \caption{{Solid lines represent the stream lines of outflow solution for 
  $x^{2}_{A}=0.75,\theta_{A}=50,\psia=55,F=0.75,q=500,\xi=1$. (a) Sideview and (b) top view.
  There are two dashed circles, one near to the center at $z\sim 0.73$ represents the Aflv\'en point location and second at
  $z\sim3500$ represents the fast point location. Here, $z$ is vertical height and $x,y$ are in terms of light
  cylinder. Inset: Region close to the base is zoomed to show the location of the Alfv\'en point (dashed circle).}}
\label{lab:fig2}
\end{figure}
	
In Fig. \ref{lab:fig2}, we plot the stream lines of outflow solution for
$x^{2}_{A}=0.75,\theta_{A}=50,\psia=55,F=0.75,q=500,\xi=1$. Figures \ref{lab:fig2}(a) \& (b) are the side and top view of stream
lines of the outflow, respectively. Here $xy$ plane represent the equatorial
plane and $z$ is the vertical height from the 
equatorial plane in terms of light cylinder. Two dashed circles, one near to the base $(z\sim 0.73$ i.e., the circle in the inset of both the panels) represents the Aflv\'en point 
location. The other at $z\sim 3500$ represents the fast point location. As we discussed before,
the transfer of angular momentum from the field to the matter, changes the direction of rotation of the flow. We can also
see in Fig. \ref{lab:fig2}, that the transfer of angular momentum from field to the matter has
twisted the stream lines of the outflow.

\vspace{0.0cm}
\begin{figure}
\hspace{2.0cm}
\includegraphics[width=12cm,height=12cm]{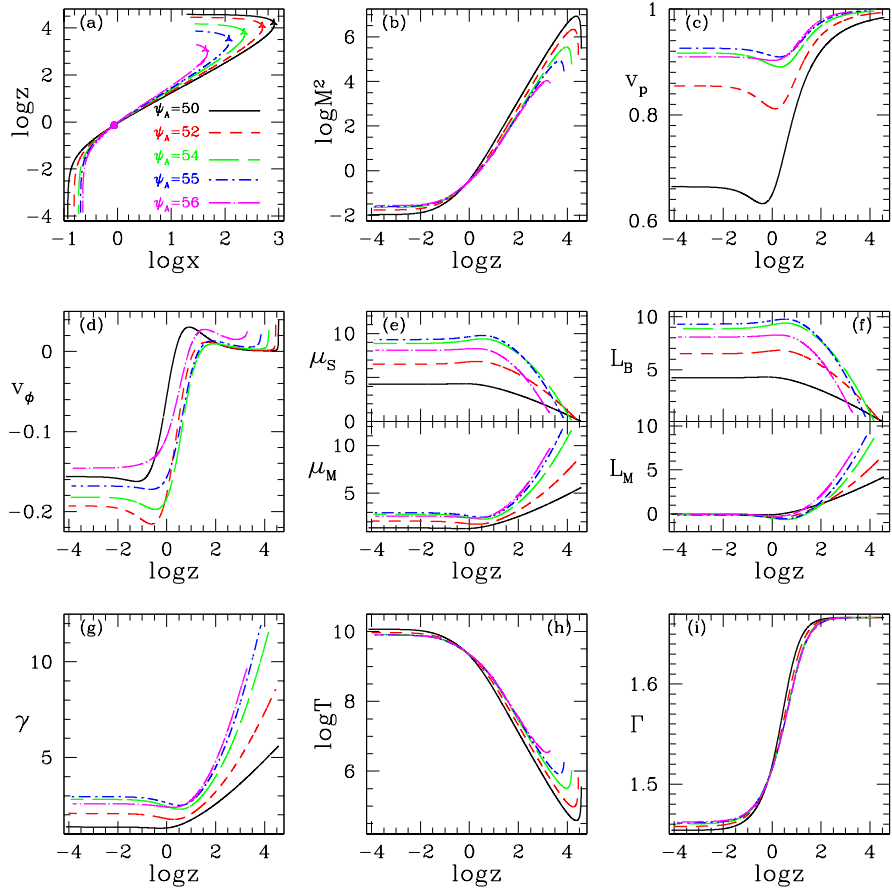} 
\caption{{Outflow solutions are for different values of $\psia=50~(\mbox{solid black})$,
$52~(\mbox{dashed red})$,
$54~(\mbox{long-dashed green})$, $55~(\mbox{dashed-dotted blue})$, $56~(\mbox{long-dashed-dotted magenta})$. 
All the curves plotted are for $x^{2}_{A}=0.75,~\theta_{A}=50,~F=0.75,~q=500,~\&~\xi=1$. Panel (a) Stream
line on the $xz$-plane, (b) $[log(M^{2})]$, (c) $\vp$, (d) $\vfi$, (e) $\mu_{S}$ and
$\mu_{M}$, (f) $L_{B}$ and $L_{M}$, (g)
$\gamma$, (h) $log(T)$ and (i) $\Gamma$ versus $log(z)$. Here, solid circles and triangles represent Alfv\'en and fast point locations.}}
\label{lab:fig3}
\end{figure}

\subsection{Solutions for different Alfv\'en point angle ($\psia$) with the disk}
In Fig.\ref{lab:fig3} we plot outflow solutions for
different values of $\psia=50~(\mbox{solid, black})$, $52~(\mbox{dashed, red})$,
$54~(\mbox{long-dashed, green})$, $55~(\mbox{dashed-dotted, blue})$ and $56$ (long-dashed-dotted, magenta).
All the curves are for fixed values of $x^{2}_{A}=0.75,\theta_{A}=50,F=0.75,q=500$ and $\xi=1$. In
Fig.\ref{lab:fig3}(a), the solution which has lower values of $\psia$ are less collimated.
Since, centrifugal force also has component in the
poloidal direction \ie $cos(\psi)$ component of centrifugal force \citep[see equation 20 in][]{vla03a},
therefore flow which has small Alfv\'en point angle with
the equatorial plane has larger centrifugal force which spreads the outflow over larger $x$.
In general, the solutions with lower $\psia$, are of lower $\mu$ and $\sigm$ and therefore are slower (i.e.,
less $\vp$). Although $\mu$ and $L$ are constants of motion, but respective magnetic and matter components of
each are not constants. The azimuthal component of velocity $\vfi$ also flips sign.
Panels Fig.\ref{lab:fig3}(h-i) show the variation of temperature and adiabatic index (varies from $1.4$ to $5/3$)
of the flow. 

\vspace{0.0cm}
\begin{figure}
\hspace{2.0cm}
\includegraphics[width=12cm,height=12cm]{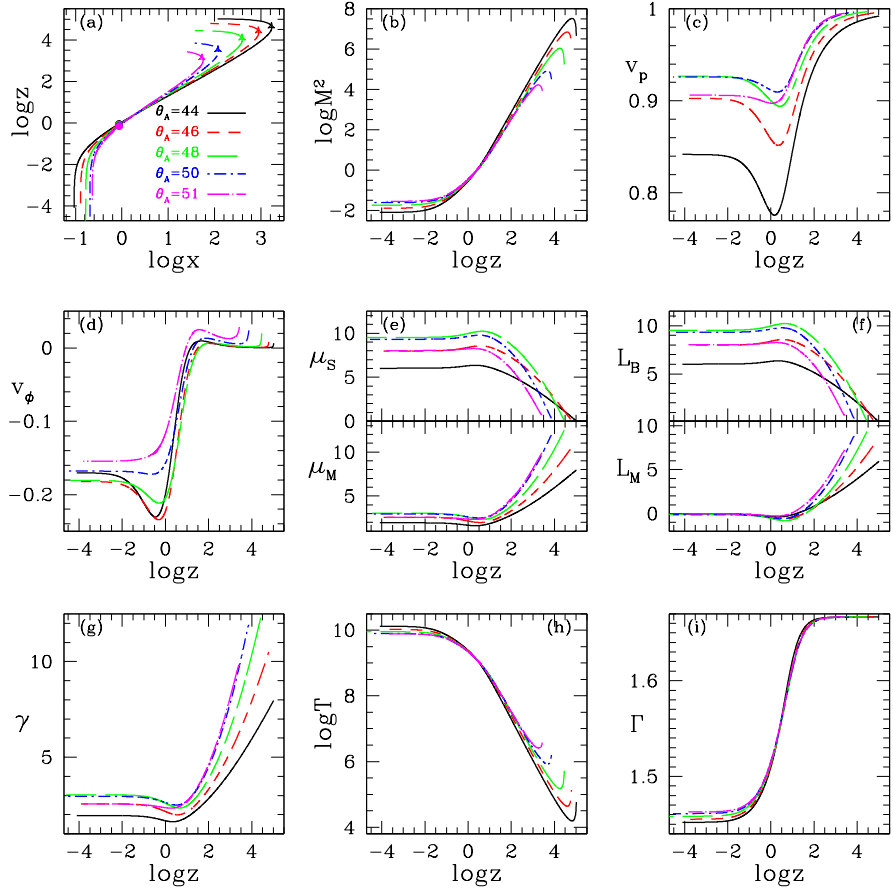} 
\caption{{Outflow solutions for different values of $\thetaa=44~(\mbox{solid black})$,
$46~(\mbox{dashed red})$,
$48~(\mbox{long-dashed green})$, $50~(\mbox{dashed-dotted blue})$, $51~(\mbox{long-dashed-dotted magenta})$ 
and four parameters are fixed \ie $x^{2}_{A}=0.75,\psia=55,F=0.75,q=500,\xi=1$ for all the curves.
Panel (a) Stream
line on the $xz$-plane, (b) $[log(M^{2})]$, (c) $\vp$, (d) $\vfi$, (e) $\mu_{S}$ and
$\mu_{M}$, (f) $L_{B}$ and $L_{M}$, (g)
$\gamma$, (h) $log(T)$ and (i) $\Gamma$ versus $log(z)$. Here, solid circles and triangles represent Alfv\'en and fast point locations.}}
\label{lab:fig4}
\end{figure}

\subsection{Solutions for different Alfv\'en point polar angle ($\thetaa$)}
In Fig.\ref{lab:fig4}, we plot outflow solutions for different values of $\thetaa=44~(\mbox{solid black})$,
$46~(\mbox{dashed red})$, $48~(\mbox{long-dashed green})$, $50~(\mbox{dashed-dotted blue})$,
$51~(\mbox{long-dashed-dotted magenta})$. Five parameters are fixed $x^{2}_{A}=0.75,~\psia=55,~F=0.75,~q=500$ and $\xi=1$ for all the curves.
Solutions with smaller $\thetaa$ start with a smaller base (small $x$), but expands to a larger $x$. While
the ones starting with larger $\thetaa$ shows exactly the opposite property. This is because the
solution with smaller $\thetaa$ have larger value of $\bfi$ near the base, but at higher $z$, $\bfi$
decreases faster than the one starting with higher values of $\thetaa$. 
In general,
$\vp$ of outflow solution is higher for higher value of $\thetaa$ ($51 \mbox{ long-dashed-dotted, magenta}$).
The $\mu_S$ and $\mu_M$ feeds at each others cost, although the total specific energy $\mu$ remains constant along the flow.
This is similar to the constancy of the total angular momentum of the flow, but components associated with the field
and the matter are not constant. 
As in the previous cases, here too the adiabatic index is not constant. 

\vspace{0.0cm}
\begin{figure}
\hspace{2.0cm}
\includegraphics[width=12cm,height=12cm]{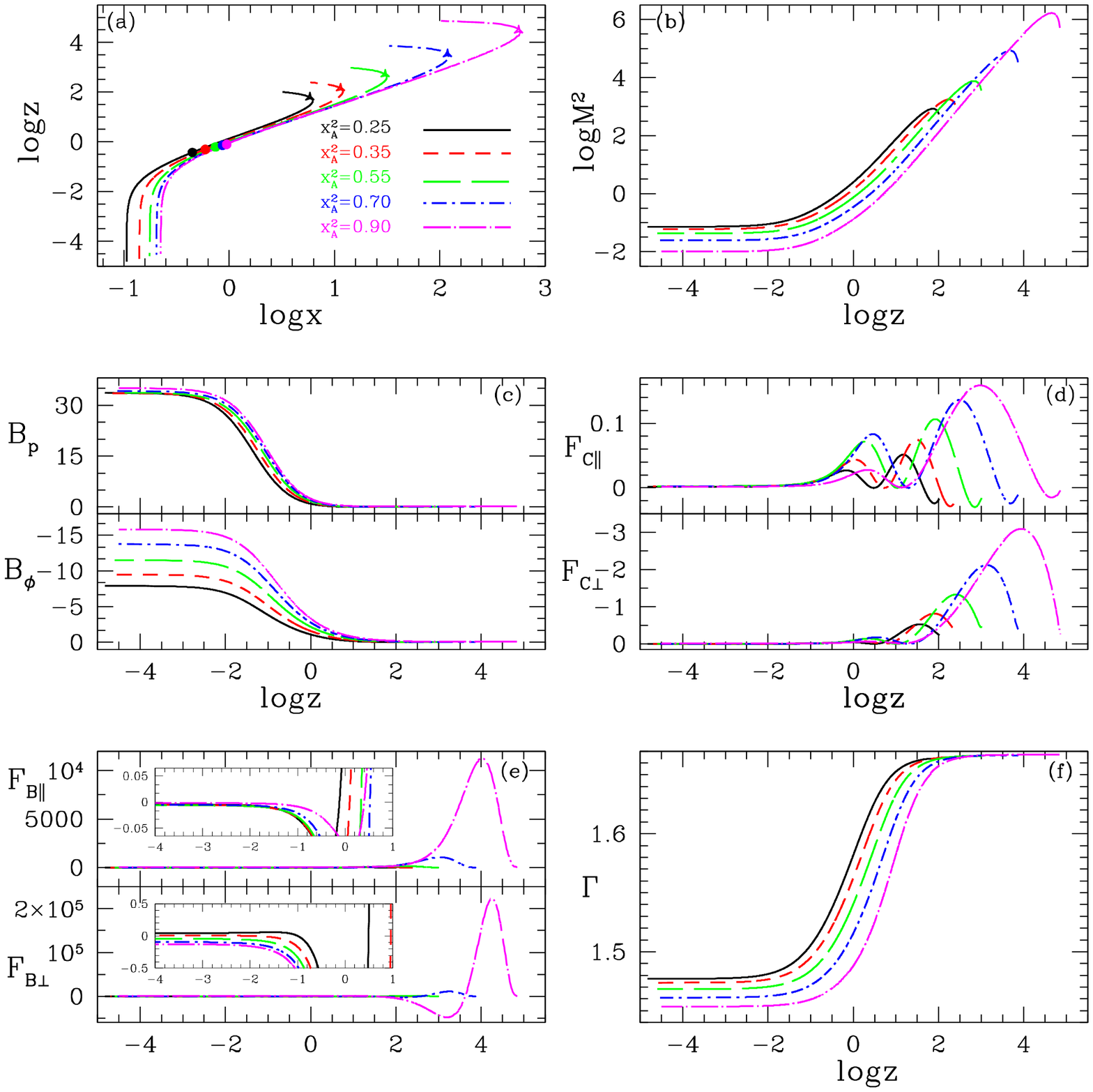}
\caption{{Outflow solutions are for different values of $x^{2}_{A}=0.25~(\mbox{solid black})$,
$0.35~(\mbox{dashed red})$,
$0.55~(\mbox{long-dashed green})$, $0.70~(\mbox{dashed-dotted blue})$, $0.90~(\mbox{long-dashed-dotted magenta})$. 
All the curves are plotted for fixed values of $\thetaa=50,~\psia=55,~F=0.75,~q=500,~\&~\xi=1$. Panel (a) Stream
line on the $xz$-plane, (b) $log(M^{2})$, (c) $\bp$ and $\bfi$, (d) $F_{{\rm C} \parallel}$ and $F_{{\rm C} \perp}$,
(e) $F_{{\rm B}\parallel}$ and $F_{{\rm B}\perp}$, (f) $\Gamma$ versus $log(z)$.
Here, solid circles and triangles in panel (a), represent Alfv\'en and fast point locations. The inset in panel
(e) zooms on to various curves corresponding to different values of $x_A$.}}
\label{lab:fig5}
\end{figure}

\subsection{Solutions for different Alfv\'en point cylindrical radius ($x_{A}$)}

In Fig. \ref{lab:fig5}, we plot outflow solutions for different values of
$x^{2}_{A}=0.25~(\mbox{solid black})$, $0.35~(\mbox{dashed red})$, $0.55~(\mbox{long-dashed green})$,
$0.70~(\mbox{dashed-dotted blue})$, $0.90$ (long-dashed-dotted, magenta).
And other parameters which are kept fixed for all the curves are $\thetaa=50,\psia=55,F=0.75,q=500$ and $\xi=1$.
The poloidal ($\bp$) as well as toroidal magnetic
($\bfi$) fields are higher for flows of higher $x_A$. However at larger $z$,
both the components of the magnetic field fall faster, compared to that in the flows of lower $x_A$ (see Fig. \ref{lab:fig5}c).
Moreover, the component of centrifugal and magnetic
forces along the streamline ($F_{{\rm C}\parallel}~\&~F_{{\rm B}\parallel}$) are larger for higher values of $x_A$.
On the other hand, collimation is achieved due to the competition between the components of magnetic ($F_{{\rm B}\perp}$)
and centrifugal ($F_{{\rm C}\perp}$) forces orthogonal to the streamline 
(Fig. \ref{lab:fig5}a, d, e). As a result, solutions corresponding to lower values of $x_A$ are more collimated (Figs.\ref{lab:fig5}a),
because the resultant of magnetic and centrifugal forces are directed towards the axis, closer to the base than those with larger
values of $x_A$. This is expected due to the assumption of radial self-symmetry.  
The $\Gamma$ distribution along the streamline for different values of $x_A$, varies significantly from each other (Fig. \ref{lab:fig5}f). It may be noted that, in almost all the cases, the outflow crosses the light cylinder with impunity. 

\subsection{Comparison of solutions for fixed and variable adiabatic index EoS (CR EoS)} 
In this section, we compared solutions of fixed adiabatic index EoS (with $\Gamma=5/3$ and $4/3$)
and CR EoS.
\begin{figure}
\hspace{2.0cm}
\includegraphics[width=11cm,height=11cm]{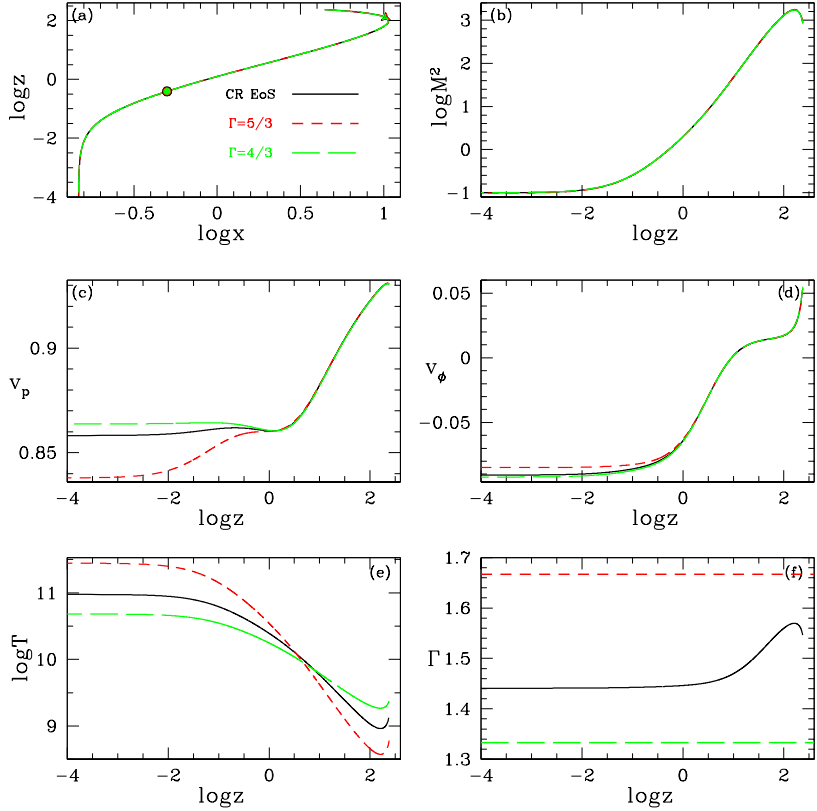}
\caption{Outflow solutions with variable adiabatic index CR EoS (solid black) with $\xi=1$, fixed adiabatic
index EoS with
$\Gamma=5/3~(\mbox{dashed red})$, and $\Gamma=4/3~(\mbox{long-dashed green})$. All curves are plotted for
$\mu=2.82420$, $x^2_A=0.25,~ \thetaa=52,~\psia=55,$ and $F=0.8$. Panel (a) Stream line on the $xz$-plane,
(b) $logM^{2}$, (c) $\vp$, (d) $\vfi$, (e) $logT$, (f) $\Gamma$ versus $log(z)$.}
\label{lab:fig6}
\end{figure}

In Fig. \ref{lab:fig6}, we plot outflow solutions for variable adiabatic index EoS or
CR EoS (solid black) with $\xi=1$ and fixed adiabatic index EoS with
$\Gamma=5/3~(\mbox{dashed red})$ and $\Gamma=4/3~(\mbox{long-dashed green})$.
All curves are plotted for $\mu=2.82420$, $x^2_A=0.25,~ \thetaa=52,~\psia=55,$
and $F=0.8$. Panel (a) shows the stream line on the $xz$-plane, (b) $logM^{2}$, (c) $\vp$, (d) $\vfi$,
(e) $logT$, (f) $\Gamma$ versus $log(z)$. In Fig. \ref{lab:fig6}a,
the streamlines of all the outflow solutions for different EoS are same. Interestingly,
all the solutions also pass through both Alfv\'en and fast critical points.
These solutions also have almost similar Alf\'en Mach number distribution (Fig. \ref{lab:fig6}b).
However, in Fig. \ref{lab:fig6}c, we can see that there is significant difference in the poloidal 
velocity and these solutions also have different values of azimuthal velocity (Fig. \ref{lab:fig6}d). 
The solutions using CR EoS, cannot be scaled with any particular fixed value of $\Gamma$. This has been
shown in many paper in the hydrodynamic (radiation hydrodynamic) limit \citep{cr09,ck16,kc17,vc19}.
As is expected, solutions of different EoS have different overall temperature
variation (Fig. \ref{lab:fig6}e). 
In Fig. \ref{lab:fig6}f, we present the variation of adiabatic index
for CR EoS and the comparison with fixed adiabatic index. For solutions with different EoS,
$T(r)$ crosses each other at some distance and yet, $\Gamma$ computed from CR EoS,
is neither $5/3$ nor $4/3$. It is clear by comparing Figs. \ref{lab:fig6}(e) and (f), that,
the temperature obtained by using $\Gamma=4/3$ is less than that obtained by using $\Gamma=5/3$,
which clearly should not be the case. 
Since, only very hot plasma should be described by $\Gamma=4/3$
and cold plasma ($T<10^7$K, i.e., $T<< \me c^2/k$) should be described by $\Gamma=5/3$, therefore, relativistic flows
described by fixed $\Gamma$ EoS clearly has a temperature discrepancy.
 
\vspace{0.0cm}
\begin{figure}
\hspace{2.0cm}
\includegraphics[width=11cm,height=11cm]{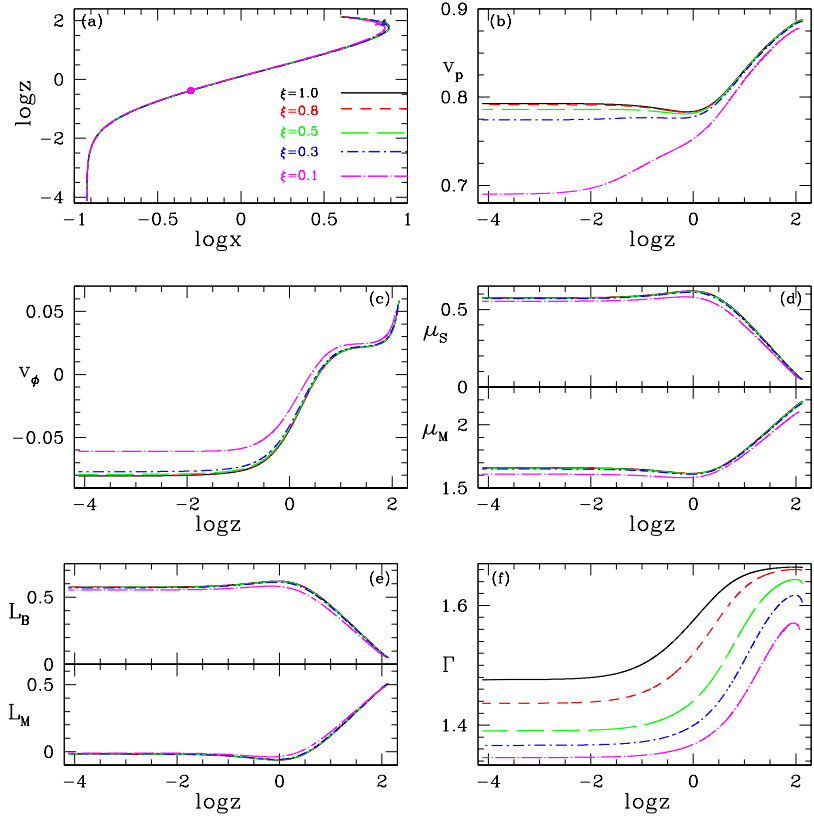}
\caption{{Outflow solutions for different values of $\xi=1.0~(\mbox{solid, black})$,
$0.8~(\mbox{dashed, red})$,
$0.5~(\mbox{long-dashed, green})$, $0.3 ~(\mbox{dashed-dotted, blue})$,
$0.1~(\mbox{long-dashed-dotted, magenta})$. All the curves are plotted for
$x^2_A=0.25,~ \thetaa=50,~\psia=55, ~F=0.75,$ and $q=500$. Panel (a) Stream
line on the $xz$-plane, (b) $\vp$, (c) $\vfi$, (d) $\mu_{S}$ and
$\mu_{M}$, (e) $L_{B}$ and $L_{M}$, (f) $\Gamma$ versus $log(z)$. Here, solid circles and triangles represent Alfv\'en and fast point locations.}}
\label{lab:fig7}
\end{figure}

\vspace{0.0cm}
\begin{figure}
\hspace{2.0cm}
\includegraphics[width=11cm,height=11cm]{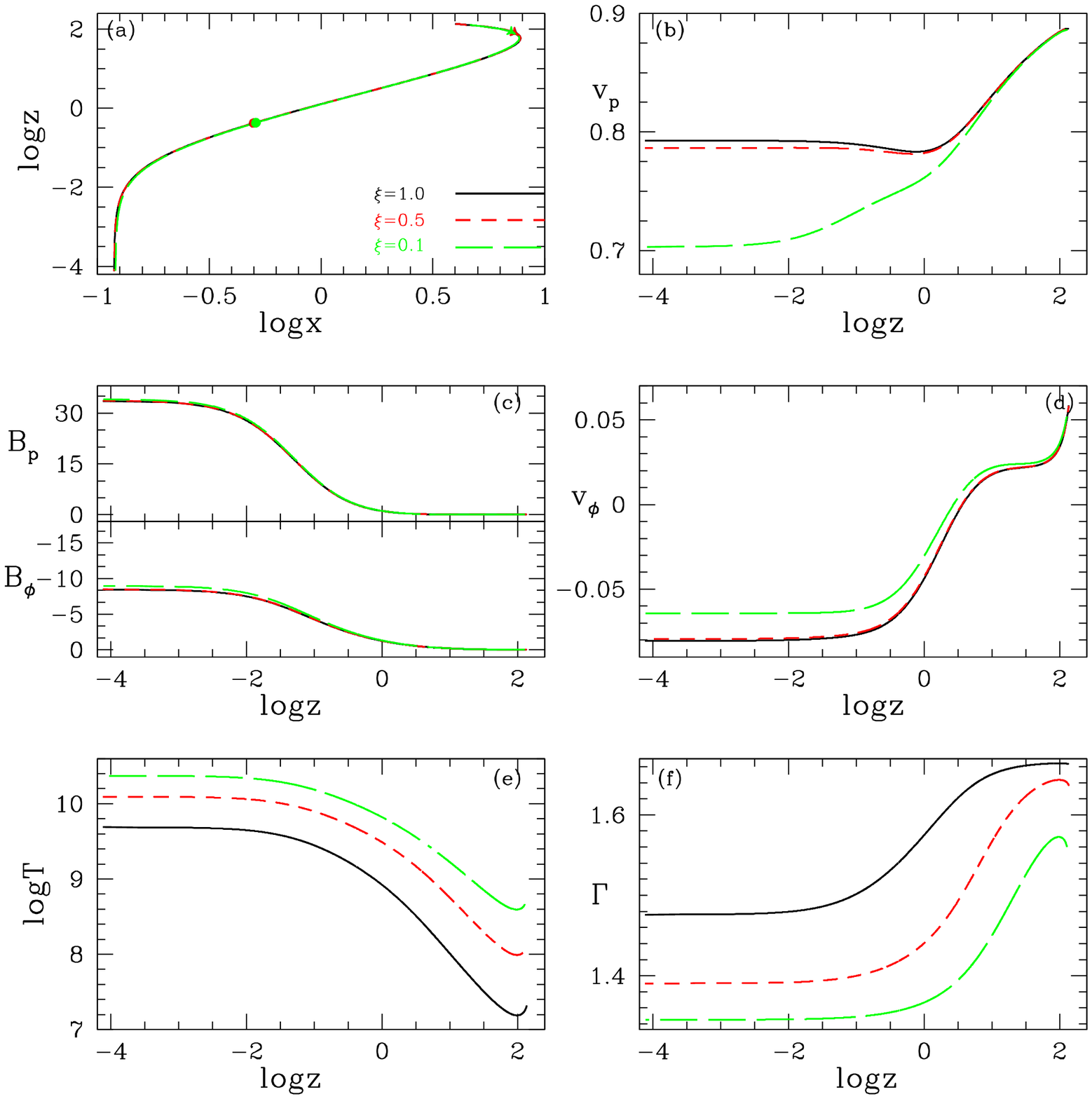}
\caption{{Outflow solutions for different values of $\xi=1.0~(\mbox{solid, black})$,
$0.5~(\mbox{dashed, red})$ and
$0.1~(\mbox{long-dashed, green})$. All the curves are plotted for
$\mu=2.23362,~ \thetaa=50,~\psia=55, ~F=0.75,$ and $q=500$. Panel (a) Stream
line on the $xz$-plane, (b) $\vp$, (c) $\bp$ and $\bfi$, (d) $\vfi$, (e) $log(T)$, (f) $\Gamma$ versus $log(z)$.}}
\label{lab:fig8}
\end{figure}

\begin{figure}
\hspace{2.0cm}
\includegraphics[width=11cm,height=11cm]{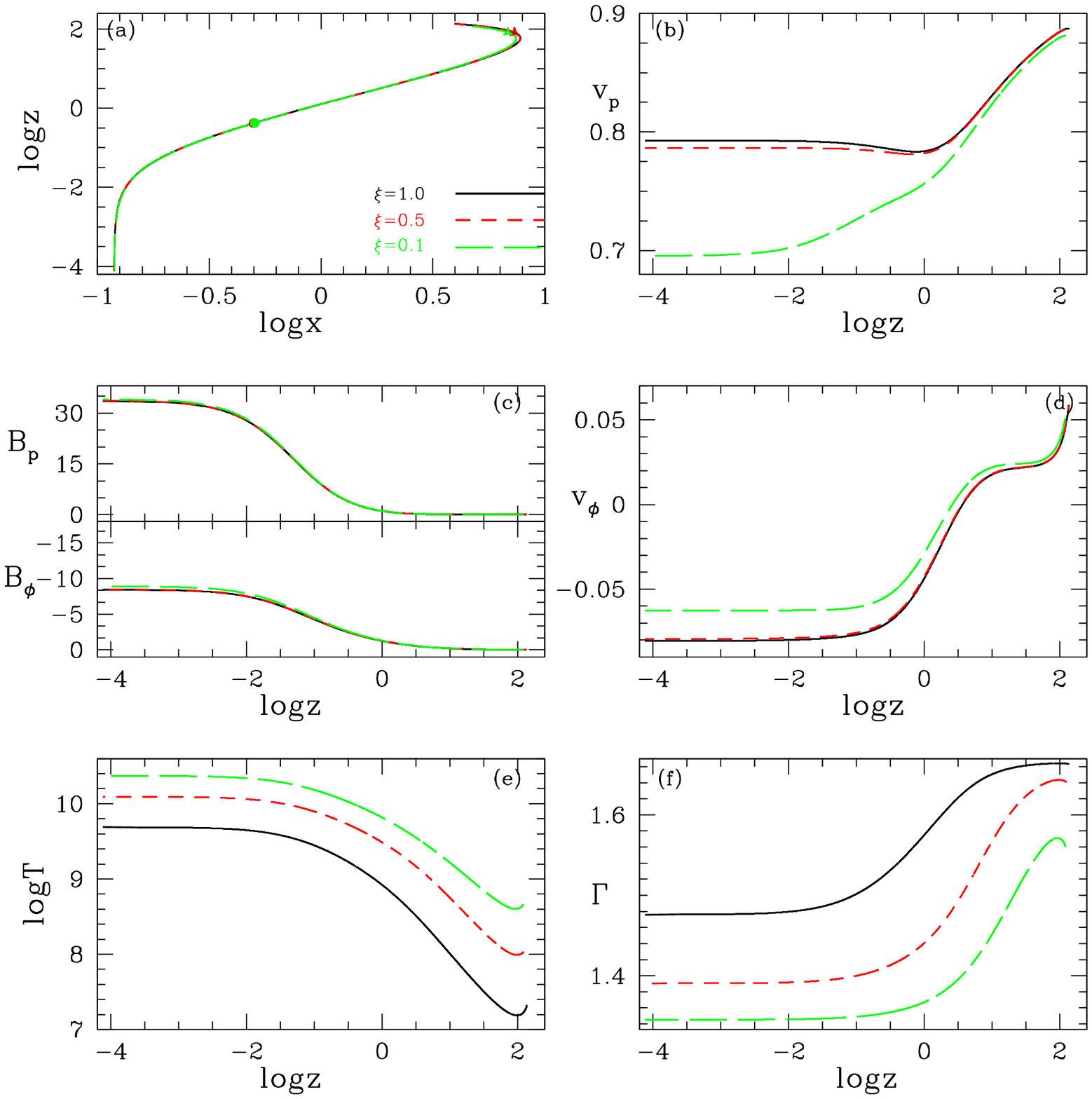}
\caption{{Outflow solutions for different values of $\xi=1.0~(\mbox{solid, black})$,
$0.5~(\mbox{dashed, red})$,
$0.1~(\mbox{long-dashed, green})$. All the curves are plotted for
$L=0.55585,~ \thetaa=50,~\psia=55, ~F=0.75,$ and $q=500$. Panel (a) Stream
line on the $xz$-plane, (b) $\vp$, (c) $\bp$ and $\bfi$, (d) $\vfi$, (e) $log(T)$, (f) $\Gamma$ versus $log(z)$.}}
\label{lab:fig9}
\end{figure}

\subsection{Solutions for different plasma compositions ($\xi$)}

In Fig.\ref{lab:fig7} we have presented outflow solutions for different
compositions, $\xi=1.0(\mbox{solid black})$ is electron-proton, $0.8(\mbox{dashed red})$, $0.5(\mbox{long-dashed green})$, 
$0.3~(\mbox{dashed-dotted blue})$, $0.1$ (long-dashed-dotted, magenta)
and other five parameters are fixed \ie $x^{2}_{A}=0.25,\thetaa=50,\psia=55,F=0.75,q=500$. 
In these solutions $\mu$ and $\sigm$ increases slightly with the increase in $\xi$, if $x_A,~\thetaa,~\psia$ and $q$ are kept constant.
It is also reflected in the plots of $\mu_S$ and $\mu_M$, as well as $L_B$ and $L_M$ (Fig.\ref{lab:fig7}e, f).
There is very little difference in the streamlines of the jets (Fig.\ref{lab:fig7}a).
However, by varying the composition of the flow from electron-proton plasma (\ie $\xi=1.0$) to
pair dominated flow $\xi=0.1$, $\vp$ and $v_\phi$ of the flow varies significantly with $\xi$ (Figs.\ref{lab:fig7} b \& c).
Even $\mu_S$, $\mu_M$ and $L_S$,$L_M$ also depend on $\xi$ (Fig. \ref{lab:fig7}d \& e).
Since $\xi$ also
influences the thermodynamics of the flow, the temperature of the jet is also crucially influenced
by its composition. As a result the adiabatic index $\Gamma$ also depends on $\xi$ (Fig.\ref{lab:fig7}f).
It may be noted, that the temperature of pair-dominated flow is
higher than electron-proton flow and therefore $\Gamma$ at any given $z$ is lower for flows with lower value of $\xi$.
Since we are comparing flows with same $x_A$ (equivalently, $M_A$), therefore
from equation \ref{amach.eq}, it can be easily shown that the temperature of pair dominated flow will be higher.
 
In Figs. \ref{lab:fig8}, we plot magnetized outflow solutions for different compositions like $\xi=1.0$ (solid, black), $0.5$
(dashed, red) and $0.1$ (long-dashed, green), but are for the same $\mu=2.23362,~ \thetaa=50,~\psia=55, ~F=0.75,$ and $q=500$.
So all these solutions are for the same Bernoulli parameter $\mu$. Since all other parameters are same, the magnetic field components
and streamlines for each are almost the same (Figs. \ref{lab:fig8}a \& c), yet $\vp$ \& $v_\phi$ (Figs. \ref{lab:fig8}b \& d)
distribution are completely different
for flows with different $\xi$. Moreover, even the temperature ($T$) and $\Gamma$ also depend on the composition
parameter (Fig. \ref{lab:fig8}e \& f). The baryon poor outflows which have same Bernoulli parameter, are slower and hotter,
compared to electron-proton flow. However, the gain in $\vp$ is more for pair dominated flow than the electron-proton flow. 

In Figs. \ref{lab:fig9}, we plot magnetized outflow solutions for different compositions like $\xi=1.0$ (solid, black),
$0.5$ (dashed, red) and $0.1$ (long-dashed, green), but are for the same $L=0.55585,~ \thetaa=50,~\psia=55, ~F=0.75,$ and
$q=500$, i. e., we compare outflows launched with the same total angular momentum (or $L$)
but different $\xi$. The streamlines are again almost the same (Figs. \ref{lab:fig9}a), however, $\vp$, $v_\phi$,
and $T$ or $\Gamma$ (Figs. \ref{lab:fig9}b---f) are significantly different for flows with different $\xi$. 

It may be remembered that the general expression of constants of motion $\mu$ and $L$ in physical units are \citep{vla03a}
$$
\mu=h\gamma-\frac{\varpi \Omega \bfi}{\Psia c^2};~L=\varpi \gamma h v_\phi - \frac{\varpi \bfi}{\Psia}
$$  
From equation \ref{enthp.eq}, it is also clear that $h$ depends on composition parameter $\xi$. So, for a given $\mu$ or $L$, if
$\bfi$ is somewhat similar at the base, then $\gamma$ (\ie $\vp,~v_\phi$) and $\Theta$ will depend on $\xi$.
That is exactly what we see in Figs. \ref{lab:fig8} \& \ref{lab:fig9}.
Dependence of flow velocity and temperature on the composition of the flow,
has also been shown in the hydrodynamic regime recently \citep{cr09,ck16,vc19,sc19}.
Therefore, it is expected that some imprint of the flow composition should be there in radiative output of the flow. 

\begin{figure}
\hspace{2.0cm}
\includegraphics[width=11cm,height=11cm]{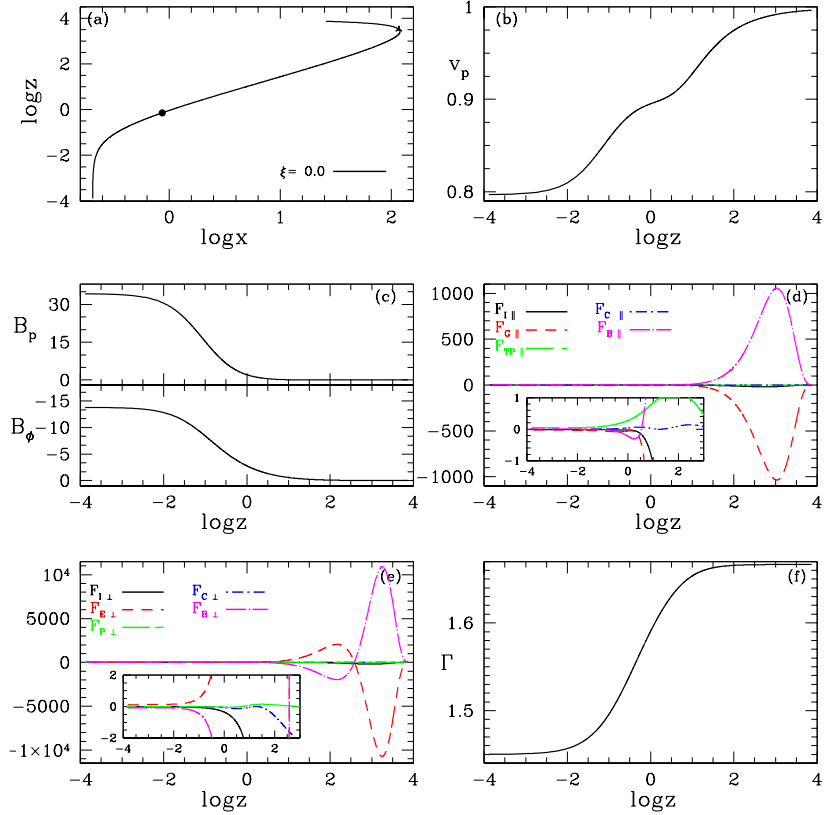}
\caption{Outflow solutions for composition $\xi=0.0$. All the curves are plotted for $x^2_A=0.75,~ \thetaa=50,~\psia=55, ~F=0.75,$ and $q=0.05$. Panel (a) Stream
line on the $xz$-plane, (b) $\vp$, (c) $\bp$ and $\bfi$, (d) parallel forces, (e) perpendicular forces, (f) $\Gamma$ versus $log(z)$.}
\label{lab:fig10}
\end{figure}

In Fig. \ref{lab:fig10}, we plot an electron-positron outflow solution or
flow having $\xi=0.0$. Other parameters are $x^2_A=0.75,~ \thetaa=50,~\psia=55, ~F=0.75,$ 
and $q=0.05$. From Fig. \ref{lab:fig10}b, it is clear that pure leptonic flow is also a trans-fast flow and the 
velocity nature is similar to proton poor flows as plotted in Fig. \ref{lab:fig8}b.
In Fig. \ref{lab:fig10}d, we plot the forces which control the poloidal 
acceleration of the flow, for example, parallel inertial force $F_{I\parallel}$, parallel `gamma' force
$F_{G\parallel}\equiv F_{GP\parallel} + F_{G\phi\parallel}$, parallel total thermal gradient force
$F_{TP\parallel}\equiv F_{T\parallel} + F_{P\parallel}$, parallel centrifugal force $F_{C\parallel}$, and
parallel magnetic force $F_{B\parallel}$ \citep[for more details see section 2.2 in][]{vla03a}. In the inset of Fig. \ref{lab:fig10}d, we can note that
these forces are comparable to each other at lower value of $z$, however for greater value of $z$, $F_{B\parallel}$ and $F_{G
\parallel}$ forces are controlling the poloidal acceleration. In Fig. \ref{lab:fig10}e, we plot all
forces perpendicular to the poloidal fieldlines, e. g., $F_{I\perp}$ (inertial), $F_{E\perp}$ (electric),
$F_{P\perp}$ (pressure gradient), $F_{C\perp}$ (centrifugal), and $F_{B\perp}$ (magnetic). Perpendicular forces have similar nature to
parallel forces, however, at a larger distances, $F_{E\perp}$ and $F_{B\perp}$ controls the collimation of
the flow. In Fig. \ref{lab:fig10}f, the adiabatic index for pure lepton flow varies from $\sim 1.44$
to $\sim 5/3$.

\section{Discussion and Concluding Remarks}
\label{sec:conclude}

In this paper we have solved the relativistic magneto-hydrodynamic equations using relativistic equation of state, in order to
study relativistic outflows.  A flow is relativistic on account of its bulk velocity (i.e.,  $\vp \lsim c$) and also
in terms of its temperature i.e., when $kT_i \gsim m_ic^2$ (subscript $i$ represents the type of constituent particle).
The first condition arises for outflows, far away from a black hole, but the second
condition especially arises in the region close to a black hole horizon which acts
as the base of an astrophysical jet. A form of EoS (CR) of a flow which can transit between relativistic to non-relativistic
temperatures has been used in this paper. As has been discussed through out this paper,
$\Gamma$ is a function of temperature in CR EoS and is automatically determined from temperature distribution.
There are a few papers in hydrodynamic regime (read in absence of ordered magnetic
field) which discusses the application of relativistic EoS in accretion and jets \citep{ck16,kc17,vc19}.
However, as far as we know, there have been no such previous attempts to solve relativistic, trans-Alfv\'enic, trans-magneto sonic 
plasma expressed by relativistic EoS and study the effect of different compositions of the plasma.
Since MHD equations are only applicable for fully ionized 
plasma, therefore, the composition of the flow is likely to either be electron-proton ($\xi=1$) plasma or 
electron-positron-proton ($0<\xi<1$) plasma.
In this paper, we have studied how various parameters like the Bernoulli constant, current distribution, the location of the
Alfv\'en point etc affect the outflow solution but only for electron-proton plasma. And then studied the effect of different EoS
and different compositions on outflow solutions.

We investigated the
contribution played by all the flow parameters, information of which shapes the final solution
of the outflow. 
We found that the current distribution affects the stream line structure, as well as the flow velocities, especially close to the base.
We also found that, not only the current distribution, the angle of the poloidal fieldline makes with the equatorial plane also affect the
solutions. In particular, the streamlines which are more inclined
to the equatorial plane are slower and less collimated. In addition, narrower the polar angle of the Alfv\'en point with the axis of the flow,
slower and less collimated is the outflow. These two angles, namely $\psia$ and $\thetaa$ are independent of each other. For a given
composition, the location of the Alfv\'en point has significant effect on the Bernoulli parameter $\mu$, the streamline and the Lorentz
factor of the flow.
We found that while the $q$ parameter which depends on the entropy, itself do not explicitly
affect the outflow solutions significantly except the temperature, but in conjunction with other parameters plays an important role. 

We have also compared the outflow solutions using fixed adiabatic index EoS, with
the one using CR EoS for a given value of $\mu$, $x_A$, $\theta_A$, $\Psi_A$ and $F$.
Although the streamlines are similar, but distribution of flow variables ($\vp$, $\vfi$, and $T$)
are significantly different. Interestingly, solutions of all the EoS, are passed through both the 
critical points (Alfv\'en and fast magnetosonic). It
may be noted, that \citet{vla03a,vla03b} only obtained trans-Alfv\'enic outflow using $\Gamma=4/3$,
but \citet{pol10} obtained trans-Alfv\'enic, trans-fast outflow solutions using $\Gamma=5/3$.
However, we showed that even with $\Gamma=4/3$,
one can obtain trans-Alfv\'enic, trans-fast outflow solution (Fig. \ref{lab:fig6}a). It appears that, depending
on the values of other parameters, there exists a critical value of $F$, below which the flow passes through both the critical points,
but for higher values of $F$, the outflow is only
trans-Alfv\'enic in nature. For example for the parameters related to Fig. \ref{lab:fig1}, trans-Alfv\'enic, trans-fast outflow is possible
if $F<0.82$.

We showed that, jets of all composition passes through the Alfv\'en and the fast point,
and get collimated to the axis after crossing the fast point. We compared solutions with different composition,
but for the same values of the Alfv\'en point, or the Bernoulli constant, or the total angular momentum. In all the cases,
composition has little effect on the streamlines, but $\vp$, $\vfi$ and $T$ distributions are significantly different.
It means that the electro-magnetic output of such outflows should also depend on the composition.
Since pair-plasma have been regularly invoked as the composition of jets,
we have also presented one case of pure pair plasma (i. e., $\xi=0.0$) outflow
solution and it nicely passes through the both critical points. The pair plasma outflow accelerates mainly in the
sub-Alfv\'enic region to super-fast region.
 The effect of composition is quite pronounced in presence of gravity
as was seen in the hydrodynamic limit \citep{cr09,kscc13,ck16} as well as, in the non-relativistic MHD regime \citep{sc18,sc19}.
So we expect the effect of CR EoS will be more 
pronounced in the RMHD limit, if gravity is considered. However, presently consideration of gravity is beyond the 
scope of this paper. It may be noted that RMHD equations combined with pseudo-Newtonian gravity have been
used to study outflows previously, with very interesting results \citep{pmm13,pmm14,cec18}.
In this paper, the jet only passes through two critical points (Alfv\'en and fast) and not the slow. The slow
point appears in presence of gravity. The existence of slow-magnetosonic point ensures low velocity and high temperature
at the base of jet, or in other words, corrects the boundary condition at the jet base. 

In all the solutions, the jet stream lines show that
there is a possibility that after crossing the 
fast point, over
collimation/magnetic field pinching can produce shock. Since the flow is moving with super-fast speed, so 
formation of shock is not going to affect the flow in the upstream and this shock location can be related to the 
fast point location. In case of M87, \citet{asa12} showed that jet radius versus jet height nicely fit
parabolic curve up to $5\times 10^{5}r_{g}$ height and after this jet radius versus height follow conical
structure. There is a dip in jet radius near the HST-1 which is located at $5\times 10^{5}r_{g}$ \ie jet radius 
versus height departs from parabolic structure and this may be due to collimation shock.

\section*{Acknowledgement}
The authors acknowledge the anonymous referee for helpful suggestions to improve the quality of the paper.

\appendix
\section{Equations of motion}
\label{appA}
The Bernoulli equation ($\mu=\gamma h- \varpi \Omega \bfi/\Psi_A c^2$) is obtained from the identity
$\gamma^{2}\left(1-v^{2}_{\phi}/c^{2}\right)=1+\gamma^{2}v^{2}_{p}/c^{2}$ (for more details see \cite{vla03a}),
\begin{equation}
\mu^{2}=\left(h^{2}+\frac{\f2\sigmm M^{4}\msin^{2}\theta}{x^{4}\mcos^{2}(\theta+\psi)}\right)\times\left(\frac{G^{4}(1-\m2-\x2)^{2}}{G^{4}(1-\m2-\xa2)^{2} - \x2 (\g2-\m2-\x2)^{2}} \right).
\label{ber.eq}
\end{equation}
Because $\mbox{tan}\psi=\frac{\partial z}{\partial \varpi}=\frac{d(G/\mbox{\small tan}\theta)}{dG}$, therefore we have,
\begin{equation}
\frac{d\g2}{d\theta}=\frac{2\g2\mcos\psi}{\msin\theta \mcos(\theta+\psi)}.
\label{dg.eq}
\end{equation} 
The transfield equation is obtained from the momentum equation by taking dot product with $-\nabla A$ then by
using equation (\ref{dengy.eq}) we can write it as,
\begin{eqnarray}
G\msin^{2}\theta\frac{d}{d\theta}\left[\mtan(\theta+\psi)\frac{1-\m2-\x2}{G}\right] &=&
(F-1)\frac{\4xa\2mu\x2}{\f2\sigmm}\left(\frac{1-\g2}{1-\m2-\x2}\right)^{2} \nonumber \\ 
& & -\msin^{2}\theta\left(\frac{\m2+F\x2-F+1}{\mcos^{2}(\theta+\psi)}\right) \nonumber \\
& & -\frac{\4xa\2mu\x2}{\f2\sigmm\m2}\left(\frac{\g2-\m2-\x2}{1-\m2-\x2}\right)^{2} \nonumber \\
& & +2\left(\frac{F-2}{\f2\sigmm}\right)\left(\frac{2\Theta h x^{4}}{K\m2}\right)
\label{transa.eq}
\end{eqnarray}
By following \cite{vla03a}, the slope of $M^{2}_A$ at the Alfv\'en point \ie $p_A=d\m2/d\theta|_{x_A}$ and
Bernoulli equation (\ref{ber.eq}) at Alfv\'en point is given by,
\begin{equation}
p_A=d\m2/d\theta|_{x_A}=\frac{2\xa2 \mcos\psia}{\siga \msin\thetaa\mcos(\thetaa + \psia)}
\label{sigA.eq}
\end{equation}
and 
\begin{equation}
\2mu=\left(h^{2}_{A}+\frac{\f2\sigmm(1-\xa2)^{2}\msin^{2}\thetaa}{\4xa\mcos^{2}(\thetaa+\psia)}\right)\times
\left(\frac{\xa2(\siga+1)^{2}}{\xa2-[\xa2-\siga(1-\xa2)]^{2}}\right)
\label{bera.eq}
\end{equation}
The Alfv\'en point condition is derived from equations (\ref{transa.eq}) and (\ref{bera.eq}) (see \cite{vla03a}),
\begin{eqnarray}
\frac{\f2\sigmm(1-\xa2)(\siga+1)^{2}}{\2mu\mcos(\thetaa+\psia)}
\bigg[-2(F-2)\frac{2\Theta_{A}}{K}\frac{(1-\xa2)}{h_{A}\xa2}\msin\thetaa \nonumber \\
+\frac{2\mcos\psia\msin(\thetaa+\psia)(\siga+1)}{\siga}+\frac{\msin\thetaa}{\xa2}\left[(1-\xa2)(F-1)-1\right]\bigg] &=& \left[\xa2-(1-\xa2)\siga\right]^{2}\nonumber \\
-(F-1)\siga^{2}(1-\xa2)-2\left(\frac{F-2}{h_{A}}\right)\left(\frac{2\Theta_{A}}{K}\right)\left(\xa2-\left[\xa2-\siga(1-\xa2)\right]^{2}\right)
\label{arc.eq}
\end{eqnarray} 
The coefficients of equation (\ref{dengy.eq}) are,
\begin{eqnarray}
A_{1}=\frac{\2mu x^{6}_{A}}{\f2\sigmm}\left(\frac{\m2}{\g2}\right)\frac{(1-\g2)^{2}}{(1-\m2-\x2)^{3}}
\frac{\mcos^{3}(\theta+\psi)}{\msin^{2}\theta\msin(\theta+\psi)}+\frac{\m2}{G^{4}}\frac{\mcos(\theta+\psi)}
{\msin(\theta+\psi)}\nonumber \\
-\frac{x^{4}_{A}}{\f2\sigmm}\frac{h^2}{\m2}\left(\frac{2\Gamma\Theta}{hK-2\Gamma\Theta}\right)
\frac{\mcos^{3}(\theta+\psi)}{\msin^{2}\theta\msin(\theta+\psi)},
\end{eqnarray}

\begin{eqnarray}
B_{1}&=&\frac{M^{4}}{G^{4}}
\end{eqnarray}

\begin{eqnarray}
C_{1}&=&\frac{h^{2}\4xa}{\f2\sigmm}\frac{\mcos\psi \mcos^{2}(\theta+\psi)}{\msin^{3}\theta\msin(\theta+\psi)}
\Bigg\{\frac{\2mu G^{4}(1-\m2-\xa2)^2}{h^{2}G^{4}(1-\m2-\x2)^{2}}-1 \nonumber \\
& &+ \frac{2\x2}{1-\m2-\x2}\frac{\2mu}{h^2}
\left[X - Y\right]\Bigg\}.
\end{eqnarray}
Here
$$
X=\frac{G^{4}(1-\m2-\xa2)^{2}-\x2(\g2-\m2-\x2)^{2}}{G^{4}(1-\m2-\x2)^{2}} 
$$
$$
Y=\frac{G^{2}(\g2-\m2-\x2)(1-\m2-\x2)(1-\xa2)}{G^{4}(1-\m2-\x2)^{2}}
$$
The coefficients of transfield equation (\ref{transa.eq}) after simplification using the expressions of
$A_{1}, B_{1}$ and $C_{1}$,
\begin{eqnarray}
A_{2}=-\msin^{2}\theta\mbox{tan}(\theta+\psi),
\end{eqnarray}

\begin{eqnarray}
B_{2}=\frac{\msin^{2}\theta(1-\m2-\x2)}{\mcos^{2}(\theta+\psi)},
\end{eqnarray}

\begin{eqnarray}
C_{2}&=&-\frac{\msin^{2}\theta(1-\m2-\x2)}{\mcos^{2}(\theta+\psi)}+\msin^{2}\theta\mbox{tan}(\theta+\psi)
\left[\xa2\frac{d\g2}{d\theta}+(1-\m2-\x2)\frac{1}{G}\frac{dG}{d\theta}\right]\nonumber \\
& &+(F-1)\frac{\4xa\2mu\x2}{\f2\sigmm}\left(\frac{1-\g2}{1-\m2-\x2}\right)^{2}
-\msin^{2}\theta\frac{\m2+F\x2-F+1}{\mcos^{2}(\theta+\psi)} \nonumber \\
& &-\frac{\4xa\2mu\x2}{\f2\sigmm\m2}\left(\frac{\g2-\m2-\x2}{1-\m2-\x2}\right)^{2}+2\frac{F-2}{\f2\sigmm}
\left(\frac{2\Theta hx^{4}}{K\m2}\right).
\end{eqnarray}

{}
\end{document}